\documentclass[10pt,twocolumn,letterpaper]{article}

\usepackage[pagenumbers]{cvpr} 

\usepackage{graphicx}
\usepackage{amsmath}
\usepackage{amssymb}
\usepackage{booktabs}

\usepackage[pagebackref,breaklinks,colorlinks]{hyperref}

\usepackage[capitalize]{cleveref}
\crefname{section}{Sec.}{Secs.}
\Crefname{section}{Section}{Sections}
\Crefname{table}{Table}{Tables}
\crefname{table}{Tab.}{Tabs.}

\def\cvprPaperID{*****} 
\def\confName{CVPR}
\def\confYear{2022}

\begin{document}

\title{DualMotion: Global-to-Local Casual Motion Design for Character Animations}

\author{Yichen Peng*\\
JAIST\\
{\tt\small puckikk@gmail.com}
\and
Chunqi Zhao*\\
The University of Tokyo\\
{\tt\small shunnki.chou@gmail.com}
\and
Haoran Xie\\
JAIST\\
{\tt\small xie@jaist.ac.jp}
\and
Tsukasa Fukusato\\
The University of Tokyo\\
{\tt\small tsukasafukusato@is.s.u-tokyo.ac.jp}
\and
Kazunori Miyata\\
JAIST\\
{\tt\small miyata@jaist.ac.jp}
\and
Takeo Igarashi\\
The University of Tokyo\\
{\tt\small takeo@acm.org}
}
\maketitle

\begin{abstract}
   Animating 3D characters using motion capture data requires basic expertise and manual labor. 
To support the creativity of animation design and make it easier for common users, we present a sketch-based interface \textit{DualMotion}, with rough sketches as input for designing daily-life animations of characters, such as walking and jumping.
Our approach enables to combine global motions of lower limbs and the local motion of the upper limbs in a database by utilizing a two-stage design strategy. 
Users are allowed to design a motion by starting with drawing a rough trajectory of a body/lower limb movement in the \textit{global design stage}.
The upper limb motions are then designed by drawing several more relative motion trajectories in the \textit{local design stage}.
We conduct a user study and verify the effectiveness and convenience of the proposed system in creative activities.
\end{abstract}

\section{Introduction}
\label{sec:1}

Creativity support systems have been extensively explored for both professional and common users in computer graphics and human-computer interaction fields. In particular, the production of character animation is a common creative process in both entertainment use and industries like game and film developments, sports, and medical applications.
Creating natural character animation requires expertise and manual labor, which prohibits novice users from even casually producing some simple character animations.
In practice, motion captures are widely used in the industry to create natural character animations.
In the motion capture process, one or more motion actors who are equipped with sensors in their body parts act the desired motions.
The word ``natural'' means to convey the credibility in the animation that there is really a virtual human there.
At the same time, the cameras in the environment sample and record the locations of the equipped sensors many times per second.
In this sense, the cost of motion capture process, which includes labor, site, time, and equipment, can be expensive and time consuming.
A big volume of motion captures have been recorded and many motion capture databases have been released for open access, such as~\cite{mocap2021} and \cite{h36mpami}.
However, the reuse of existing motion capture database is still difficult for novices:
for specific applications and stylish motion details, animators still need to record motion captures from scratch. 
Although existing motion capture databases usually provide keyword-based motion search, the subjectively defined keywords can not represent all details in each motion sequence and thus prohibit animators from retrieving desired motion sub-sequences from a large scale database.

\begin{figure}[t]
	\begin{center}
	\includegraphics[width=\columnwidth]{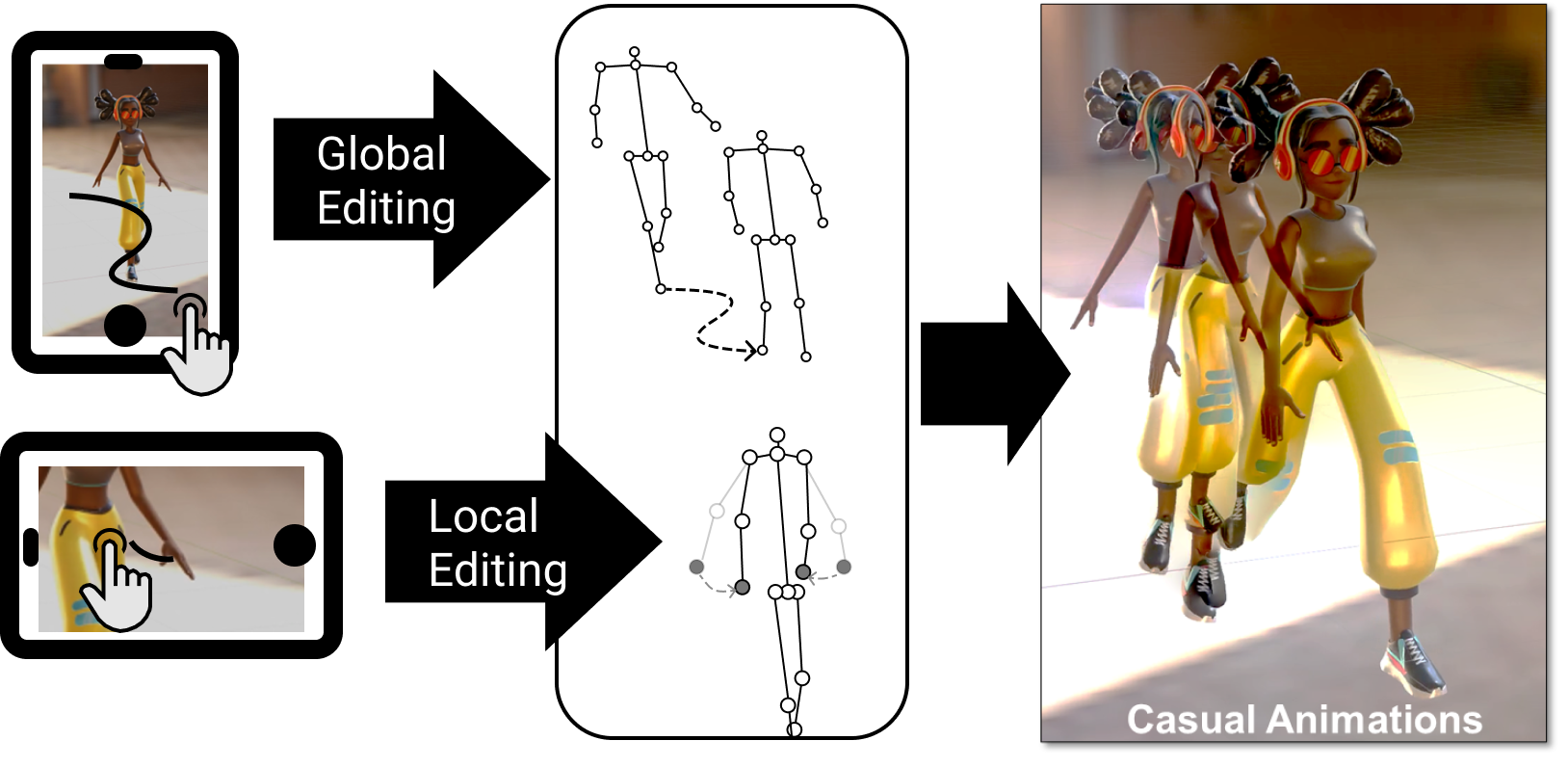}
	\end{center}
	\caption{\textit{DualMotion} enables users to input rough sketches and create casual character animations.}
	\label{fig:1}
\end{figure}

As another option for motion retrieval, sketch-based interfaces have been developed and researched by a previous work~\cite{peng2021nico}.
The sketch-based interfaces are for computer graphics applications, such as interactive animation design~\cite{hu2019sketch2vf,2014draco,xing2016energy}, 3D modeling~\cite{sketch2model2021} and mesh editing~\cite{autoscuplt}, and have been proven to be intuitive to users especially novice users.
Nevertheless, applying sketch-based interfaces in motion retrieval and editing faces some challenges. 
Motion capture takes a complicated format by its spatio-temporal nature.
A motion capture sequence consists of a \textit{static} part and a \textit{dynamic} part.
The static part defines the character's skeletal structure along with node offsets.
The dynamic part defines the character's changing pose in each frame, represented by the relative node rotations w.r.t. (with reference to) each node's parent node.
In a typical human character animation, the skeletal structure may contain over ten nodes, and the dynamic sequence can consist of thousands of frames.
On the contrary, 2D sketching space is much simpler than the motion capture data in terms of dimension and is not expressive enough for tasks such as motion retrieval and composition.
Particularly, the spread of touch screen devices and short video applications proposes a potential situation to create character animations, where the operation range and accuracy are both limited.
In such a situation, the fact that users can be ``casual'' and can only input with rough sketches makes the task ``sketch to animate characters'' even more challenging.

To solve these issues, we propose \textit{DualMotion}, a multi-stage sketching scheme for character animation retrieval and composition considering novices as target users in a casual animating situation (the operation range and accuracy are limited, while the naturalness of the animation can be sacrificed somewhat for casual creation).
The key idea is to decompose the challenging and complicated design task into several stages.
In each stage, a novice user only needs to accomplish a sub-task and pursue a sub-objective.
For the sketch-based motion retrieval and composition task, we decompose the entire design process into two stages:
The first one is the \textit{global stage}, where the reference point is the virtual ground and the user draws a rough trajectory to represent the entire movement of the character;
the second one is the \textit{local stage}, where the reference point is the character's center of mass and the user draws relative trajectories to represent the detailed movement of character limbs.
We adopt a data-driven approach to assist users in retrieving and combining desired motions from databases.
In addition, in either of the two stages, we show the retrieved motion candidates in dashed polygons as shadow guidance.
We implement the proposed method and conducted a user study to compare the two-stage sketching interface with an one-stage baseline.
Our qualitative and quantitative evaluation results showed that the proposed method can significantly (p < 0.01) assist novice users in searching and editing their desired character animations.

We summarize our main contributions as below:

\begin{enumerate}
    \item A creativity support system with two-stage character animation retrieval and composition scheme that decomposes the challenging task into stages. In each stage, novice users just need to concentrate on a sub-goal of the design process.
    \item The implementation called \textit{DualMotion},which utilizes the proposed two-stage design scheme. \textit{DualMotion} supports novice users in designing usual motion in a data-driven manner.
    \item A user study involving 14 participants to validate the effectiveness and usability of the proposed scheme and the \textit{DualMotion} implementation.
\end{enumerate}

\section{Related Work}
\label{sec:2}
This section reviews prior work on frameworks for (1) character animations authoring and (2) motion retrieval and composition. 

\subsection{Authoring Character Animations}
Some systems enable users to design character animations using physical-based simulation~\cite{xing2016energy,kazi2016motion,willett2017secondary} and users' demonstration~\cite{Dvoroznak20-SA,guay2015dynamics,guay2015space}. 
Although these methods can be suitable for simple object motions, such as graffiti, they are unsuitable for skeletal character motions. The reason is that skeletal data consists of $N$ joints ($N > 20$), that is, to make natural-looking motions, users must repeatedly set all joint angles at each frame, which might be very time consuming and tedious. 
To address this issue, animators often use motion capture data (Mocap), which are records of the actual movement of objects or people. 
\textcolor{black}{Dontcheva et al.~\cite{siggraph03} enabled the animator to author character animations by imitating the movement under a customized Mocap system.} 
Besides, the large scale Mocap databases have been constructed by many research institutions and companies~\cite{mocap2021,h36mpami}, and many works to generate motion sequences using the Mocap databases have been proposed~\cite{ho13topology,iwamoto19automatic}. 
Then, we take a ``data-driven'' approach but consider (i) how to intuitively find motion data from the database and (ii) how to edit motion data for our needs. 

\subsection{Retrieving and Editing Character Motion Data}
Against this background, retrieving motion data from large motion databases has been thoroughly investigated, for example, typing keywords.
\textcolor{black}{Kruger et al.~\cite{sca10} introduced a similarity searches algorithm for motion retrieval in a large-scale Mocap databases.}
Peng et al.~\cite{peng2021nico} introduced a sketching interface to retrieve motion data by drawing motion trajectories on a screen. Unlike static pose drawings~\cite{choi2012stick}, motion trajectories are suitable for imagining the change of character motions. However, we need to edit the retrieved data to meet the user's needs, rather than use the data as is.

Motion Doodle~\cite{thorne2004motion} segments the input sketch of motion trajectory into several primitives (e.g., walk and jump) and generates simple motion sequences (e.g., walk $\rightarrow$ jump  $\rightarrow$ jump). However, this system focuses only on the root motions of the character (located at the hip), called ``global'' motion, so it remains difficult to locally edit character motions, such as hands and legs. By contrast, Choi et al.~\cite{choi2016sketchimo} proposed SketchiMo, which allows users to directly edit trajectories of character motions. These approaches enable users to directly edit local motions one by one on the screen, but they require local coordinates of the character joints at each frame to be carefully tuned. \textcolor{black}{MotionMaster~\cite{sca06} built a database of Kungfu motion which allowed users to sketch the start and end poses of the character and draw the trajectory of a user-specified joint movement after labeling those joint pairs. However, this system required a high level of drawing skill with the time-consuming labelling tasks.} 

Building on these approaches, we simply separate the sketch process of motion trajectories into two stages: (i)~global stage (i.e., root motion retrieval) and (ii)~local motion stage (i.e., limb motion retrieval), and make motion sequences by combining the retrieved global/local motions.

\section{Global-to-Local Motion Design}
\label{sec:3}
The proposed character animation retrieval and composition scheme was inspired from the following observations:
character animation design is challenging for novice users because of the complicated data format of motion data.
In Addition, limited by their experience and skills, novice users may find the design task difficult as a single process and suffer from the blank page syndrome.
To address this issue, we propose a two-stage design scheme to assist novice users in designing motions of character animations with both global and local stages.
We decompose the whole motion design process into two stages. 
In each stage, users can just concentrate on one sub-goal of the design process.
Users begin from the global stage and sketch on the rough movement of the character's lower limb.
After users are satisfied with the work, they enter the local stage and draw the detailed movements of the character's upper limbs. In this section, we go through the proposed motion design scheme and describe the global and local stages in details.

\begin{figure}[ht]
	\begin{center}
	\includegraphics[width=1.0\columnwidth]{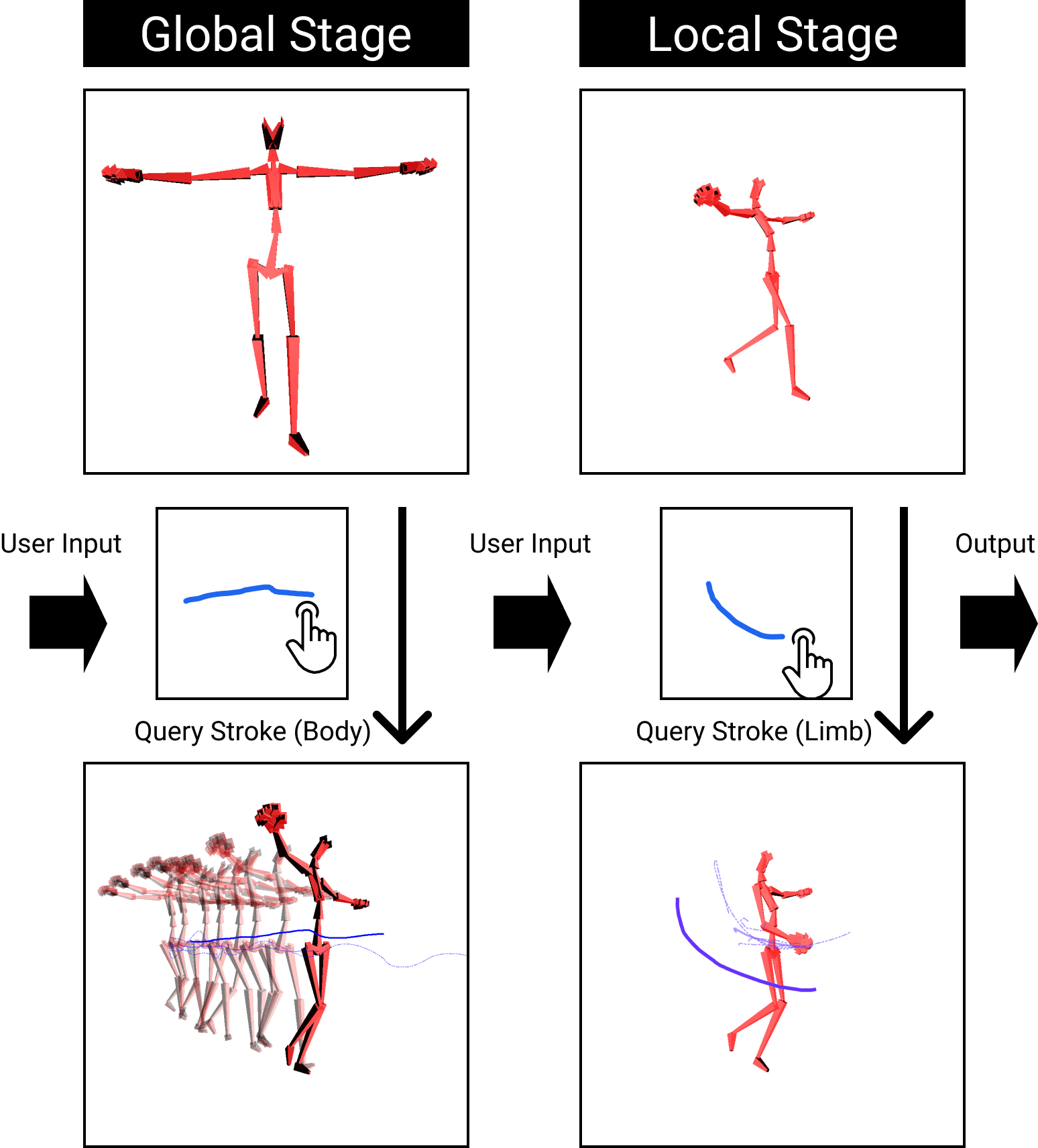}
	\end{center}
	\caption{The proposed Global-to-Local design scheme for data-driven character animation design. In the global stage, users retrieve the rough motion of the whole character by inputting a query stroke; in the local stage, users further draw the strokes to edit the upper limb motion.}
	\label{fig:2}
\end{figure}

\subsection{Overview}
In the proposed character animation authoring scheme, the character is shown on a virtual ground.
The users draw query strokes that represent motion trajectories of the character nodes.
To fully utilize and reuse existing Mocap databases, we conduct a trajectory similarity comparison between the user-input query strokes and the node trajectories in the database by projecting them into the same 2D camera coordinate system.
\textcolor{black}{Because the main purpose of the current prototype is  the locomotion retrieval, we only adopted the motions of walking, running, jumping and punch and kicking for simplification.}
The overall pipeline is shown in Figure~\ref{fig:2}.

\subsection{Global Stage (Lower Limb)}
Global stage is the first stage and the default state of the two-stage design scheme.
When users enter the stage by opening an initial scene, a blank virtual ground is displayed.
To start the query and further composition, the users draw a trajectory on the screen, to represents the ground-projected moving trajectory of the character's center of mass.
The system then starts to match the user-input query stroke with the projected trajectories in the database, and apply the best-matched animation on the character in the interface.
The users can choose the desired rough movement of the whole character and enter the local stage to specify more details of the character's upper limbs, or just repeat the retrieval in the global stage.
In the global stage, we define the virtual ground as a reference point, which allows the users to freely adjust the position and angle of the camera.

\subsection{Local Stage (Upper Limb)}
After the users are satisfied with the retrieved global movement of the character, they can continue the design process and choose a specific upper limb for further detailed retrieval.
The reference point of the local stage is the character's center of mass, which enables a surveillance camera that always follows the character and stares at the character's center of mass.
The camera locates on a sphere whose radius is adjustable and center is also the character's center of mass.
Similarly, the users draw trajectories of the selected limb node with respect to the center of mass.
When the users finish the query stroke input, the system translates all trajectories of the selected limb node from the database by projecting them to the camera coordinate system and starts similarity matching.
The users then pick the desired limb motion from the retrieved results, and the system applies the motion to the selected limb based on the rough motion retrieved in the global stage to make a combined new animation.
With iterations between global and local stages on different limbs, the users finally make a new animation from existing motion captures and save it to end the design process.
Although legs should be considered as upper limbs, we only enable head and hands motion retrieval and composition in the local stage because leg motion is highly entangled with the overall movements of the character.

\section{System Framework}
\label{sec:4}
In this section, we explain how we implement the motion editing system prototype called \textit{DualMotion} in the proposed framework.

\subsection{Dataset Construction}
To build the data-driven motion editing system, we construct the motion dataset by selecting the locomotion data from the CMU Mocap data library~\cite{mocap2021}, which consist of several motion category, such as locomotion, physical activities \& Sports, etc. (Note that our data is mainly chosen from the locomotion.) 
We empirically choose motion data with similar sizes and skeletal proportions. 
We then manually trim the motion data into 100 frames to allow them to share the same duration, and normalize all the motion sequence into a same initial position that is the origin of the virtual spatial coordinate system. 
In our case, the root node (hip) movement of each motion data is considered the global movement, and each limb movement is considered the relative local movements of the root node.
Lastly, each motion data point is split into limbs and hip, which are stored.

\subsection{Trajectory Representations}
The users directly sketch on the widget that visualizes the character animation for both global and local stages. 
That is, the sketching canvas coordinate system shares with the camera coordinate system of the visualization widget. Note that as stated in Section \ref{sec:3}, the editing tool maintains a camera with different reference points in the global and local stages.

In each stage, motion trajectories in the motion dataset are projected into the canvas coordinate system, as follows: 
\begin{equation}
    \mathbf{V}_{canvas}^{j}(i, t) = M_{proj} \mathbf{V}_{orig}^{j}(i, t)
\label{eq:1} 
\end{equation}
%
\noindent
where $\mathbf{V}_{orig}^{j}(i, t) \in \mathbb{R}^3$ is the 3D coordinate of $i$-th node at time $t \in \{1, T\}$ in the $j$-th motion data in the database, $\mathbf{V}_{canvas}^{j}(i, t) \in \mathbb{R}^2$ is the corresponding projected 2D coordinates of the motion trajectory, and $M_{proj} : \mathbb{R}^{3} \rightarrow \mathbb{R}^{2}$ is a 3D to 2D projection matrix of the maintained camera. 
In this paper, we represent the motion trajectory in the database as follows:
\begin{equation}
V_{orig}^{j}(i) = \{\mathbf{V}_{orig}^{j}(i,1), \mathbf{V}_{orig}^{j}(i,2), \cdots, \mathbf{V}_{orig}^{j}(i,T)\}
\end{equation}

\noindent
where $T$ is the number of the motion frame.
When the users pan or zoom to relocate the camera so that the projection matrix $M_{proj}$ changes, the proposed tool updates the projected trajectories $V_{canvas}^{j}(i)$ $= \{\mathbf{V}_{canvas}^{j}(i, 1), \cdots, \mathbf{V}_{canvas}^{j}(i, T)\}$ for comparing the user query sketch with the motion in database.

\begin{figure}[t]
    \centering
    \includegraphics[width=1\linewidth]{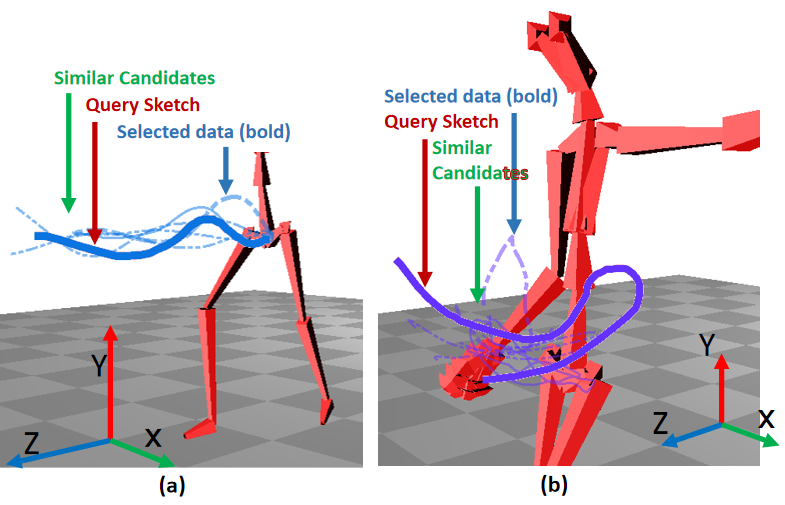}
    \caption{Shadow-like guidance that displays joint movement projected trajectory in global (left) and local (right) design stages.}
    \label{fig:shadow}
\end{figure}

\subsection{Trajectory-Based Retrieval}
Similarly, we represent a user-input stroke $S_{user}$ for both two stages as follows:
\begin{equation}
S_{user} = \{\mathbf{S}_{user}(1), \mathbf{S}_{user}(2), \cdots, \mathbf{S}_{user}(T)\}
\label{eq:2}
\end{equation}

\noindent
where ${S}_{user}(t) \in \mathbb{R}^2$ is the coordinate of $t$-th sampling point of the user-input stroke, and $T$ is the number of total sampling points of the stroke. 

We then compute a similarity between the query stroke $S_{user}$ and the projected trajectory of the $j$-th motion data in the database $V_{canvas}^{j}$, as follows: 
\begin{equation}
Sim^{\:j}(i) = F(S_{user}, V_{canvas}^{j}(i))
\label{eq:3}
\end{equation}

\noindent
where $F(a,b)$ is the Fr\'{e}chet distance between two strokes $a$ and $b$. Note that $i$ represents the root node in the global stage, and the limb nodes in the local stage. 

After computing all the similarities, we pick the top $N$ relevant retrieval results in the database ($N=5$ in our implementation) and merge them as a shadow-like guidance to display their projected trajectories in real-time, inspired by \textit{ShadowDraw}~\cite{lee2011shadow}.
As shown in Figure.~\ref{fig:shadow}, the user can click on the most desirable motion candidate for selecting global/local motion data. Benefiting from the guidance, users understand the locations and shapes of each stroke drawing.

\subsection{Motion Sequence Synthesis}
Given the retrieved global and local motions, \textit{DualMotion} generates a final motion sequence. 
In our prototype, we utilized a BVH data format, which consists of the position of the root node and all joints' rotation at each frame, and simply assigned the rotational matrices of the associated nodes in the retrieved local motion to the retrieved global motion.
\textcolor{black}{For instance, the rotational matrices of left shoulder node and its child nodes would be modified when the left hand movement is authored. Similarly, right-shoulder node, lower-neck node and their child nodes are allowed to be re-assigned when the related movement is modified.}
The similar approach is applied by Iwamoto et al.~\cite{iwamoto19automatic} which they segmented the hands movement and body motion to synthesize the dance motion sequence. Although this approach seems to work well, other methods can also be used in our framework.

\begin{figure*}[t]
  \includegraphics[width=\textwidth]{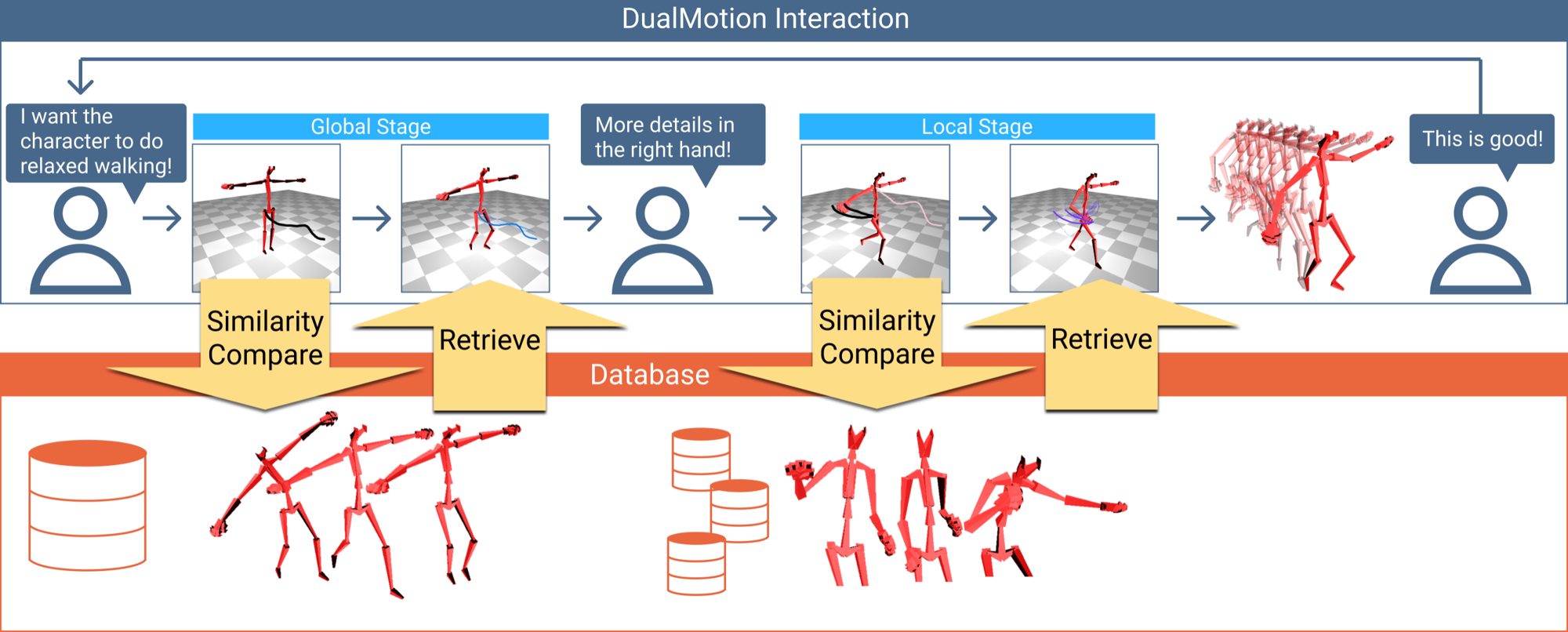}
  \caption{The design process of \textit{DualMotion}.}
  \label{fig:editing_process}
\end{figure*}

\begin{figure}[t]
    \centering
    \includegraphics[width=1.0\linewidth]{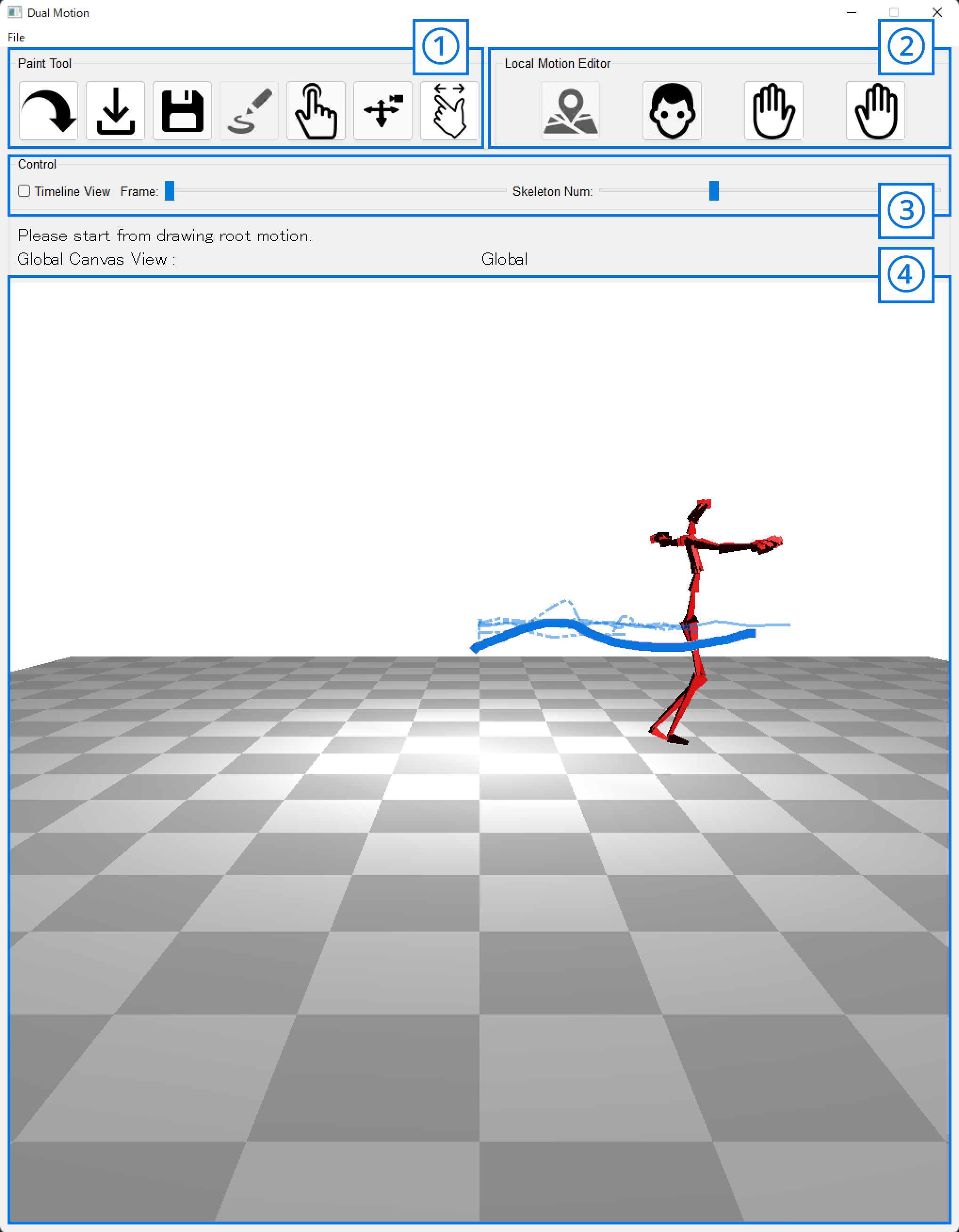}
    \caption{Interface of \textit{DualMotion}. It consists of: (1) a tool panel contains buttons like undo, load, save, draw, select, camera pan, and camera zooming; (2) a stage panel to switch between global and local stages; (3) a control panel for timeline visualization and (4) a graphical interface for drawing and character animation visualization.}
    \label{fig:interface}
\end{figure}

\subsection{User Interface}
We implement the two-stage character animation design scheme as a prototype called \textit{DualMotion} using openGL and Qt. 
Figure~\ref{fig:interface} shows the user interface of \textit{DualMotion}.
It consists of a graphical widget and a control panel.
The graphical widget is main region for users to interact (visualization and drawing).
At the top of the graphical widget, the control panel provides buttons to enable basic functions (undo, load, save, etc.), mode selection between drawing and result selecting, camera controlling, global and local stage switching, and timeline visualization.
We have introduced the the basic design scheme in \ref{sec:3} and \ref{sec:4}.
As an alternative for showing the character animation, we implemented a timeline visualization to help users check the entire animation in a static way rather than waiting for the animation playback.
The character animation is shown in the visualization widget by sampling the frames in the motion sequence, and the character skeleton of each frame is statically overlaid in the graphical widget.
Only one frame of the character skeleton is highlighted, whereas the others are semi-transparent.
A slider is used for adjusting the position of the highlighted frame, and another slider is used for manipulating the frame interval.

\subsection{Design Process}
The design process of \textit{DualMotion} is shown in Figure~\ref{fig:editing_process}.
The user starts from the global stage with a rough objective about the character animation in mind and then goes forward into the local stage and decides more details in upper limbs.
The user can repeat the two design stages until satisfied with the final output.
Every time the user inputs query strokes, the \textit{DualMotion} interface feeds forward the user inputs, compares them with the database, and render the retrieved and composited motion sequences.


\section{User Study}
\label{sec:5}
To evaluate the validation of \textit{DualMotion}, we conducted a user study that consists of a comparison study and a post-experiment questionnaire. In this section, we describe the experimental details and discuss the evaluation result.

\subsection{Comparison Study}
To evaluate the validation of the proposed user interface, we conduct a user study to compare \textit{DualMotion} with a one-stage motion retrieval UI, which has the same function as \textit{DualMotion}, except that it only allows \textcolor{black}{users to edit and synthesize motion to obtain the results but only in the global stage editing (i.e. hip movement). For example, users are allowed to edit a hand waved movement while the character is stepping forward. It is difficult for users to define such a hand movement on a moving character in this case.}
Similar to \textit{DualMotion}, the implemented one-stage UI provides the shadow guidance to aid users in finding similar results (trajectories of motion) after each time they finish drawing a stroke. 
However, in contrast to \textit{DualMotion}, it only allows users to retrieve motion by drawing the trajectories in a global manner.

The study consists of two sections: 1) motion design with a reference motion, 2) free motion design. 
In the first section, we provided five skeleton motions that consist of different identities of joint movements from the prepared database.
By using \textit{DualMotion} and the one-stage edit UI respectively, the participants were required to retrieve and edit the motion following one of these references randomly.

On the one hand, to compare the efficiency, we recorded the time cost and the number of operations (mouse clicks). 
On the other hand, to compare the quality of the design result, we also calculated the mean absolute error between the synthetic motion and the reference motion to compare the similarity, which confirms the accuracy of the user design result in the two different UIs.
In the second section, to evaluate whether our interface meets user intent, the participants are asked to do a free motion design using \textit{DualMotion} in 5 minutes.
They are asked to answer their editorial intent or the reason why they would like to design such kind of the character motion in the post-experiment questionnaire.

\subsection{System Evaluation}
Except to answer the editorial intent in free design task, the participants are required to evaluate the overall system by answering the questions rely on System Usability Scale (SUS)~\cite{SUS} metrics after the experiment.
To rate the perceived workload of design motion using \textit{DualMotion}, we requested the participants to fill in a post-experiment questionnaire designed following NASA Task Load Index~\cite{NASA} (Considering the length of the questionnaire, we utilized the \textit{raw} version of NASA-TLX in our evaluation study where weighing each of the evaluation item is not necessary).

\subsection{Experiment Process}
We invited 14 graduate students at the age of approximately 25 years to participate in our experiment. 
We choose a total of 55 skeleton motion data from the CMU~\cite{mocap2021} motion database for retrieval in the user study, which are gait data in different directions with different styles, including fast walk, slow walk, duck walk, zombie walk, and so on.
They had approximately 5 minutes to get familiar with each tool before the task.
In addition, a player window with a modifiable virtual camera is provided so that reference motion clips can be viewed at any time during design.
The order of finishing tasks using one-stage UI and \textit{DualMotion} are randomly shuffled to make sure that we equally evaluated each UI.
The free motion design experiment is the final task that we believed that the participants are familiar enough to user \textit{DualMotion} to create some motion following their intent.
Lastly, they are required to fill the post-experiment questionnaire mentioned above.

\begin{figure}[t]
    \centering
    \includegraphics[width=1\linewidth]{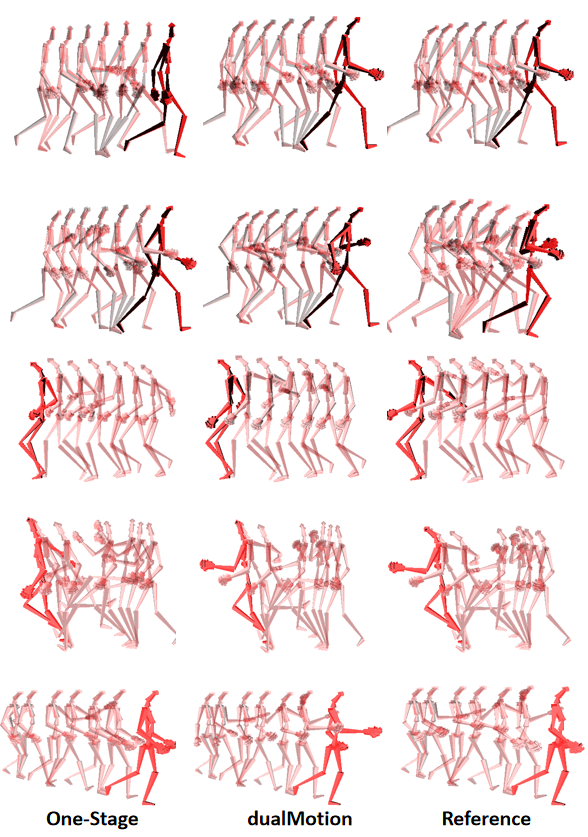}
    \caption{Examples of participant design results by using each interface for the five motion references. The design results using DualMotion are more similar to the reference compared to the one-stage interface, especially the hands movements .}
    \label{fig:comparative}
\end{figure}


\section{Results}
\label{sec:6}
\subsection{Evaluation Results}
Figure~\ref{fig:comparative} (right) shows various reference motions which used in our user study.
For the validation of DualMotion, \textcolor{black}{we compared the Euler angle of $i$-th node at time $t \in \{1, T\}$ in the user-designed results $\mathbf{x}_{i}(t)$ to the reference $\mathbf{x}'_{i}(t)$ and computed the MSE loss as follows:}

\begin{equation}
\textcolor{black}{MSE = \frac{1}{TN}\sum_{t = 1}^{T}\sum_{i = 1}^{N}\Vert \mathbf{x}_{i}(t) - \mathbf{x}'_{i}(t) \Vert^2}
\label{eq:5}
\end{equation}
\textcolor{black}{where $N$ is the number of joints in the character skeleton.} The edited motion results by using \textit{DualMotion} have a higher similarity compared with the one-stage design approach. 
In addition, we analyzed the overall MSE result of \textit{DualMotion} by running a one-tailed $t$-test. The $t$-value was 2.48, and the $p$-value was 0.01 ($p < 0.05$), which revealed significant differences in the MSE loss.

\begin{figure}[t]
    \centering
    \includegraphics[width=0.95\linewidth]{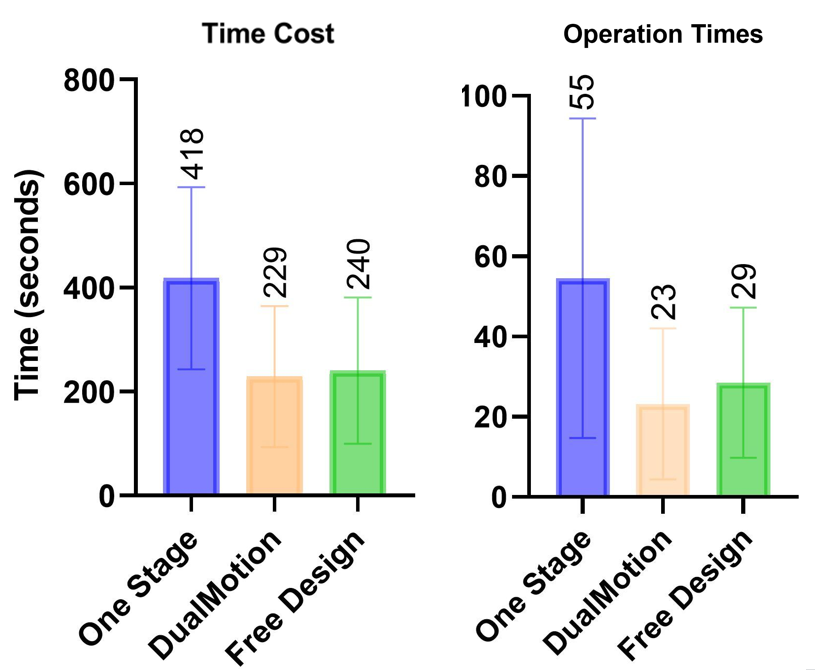}
    \caption{Average time cost (left) and operation times (right) of each task. Noted that the free design task was only done by using \textit{DualMotion}. It shows that the participant were able to design a expected motion within few minutes.}
    \label{fig:time&operation}
\end{figure}

For the evaluation of the efficiency, Figure~\ref{fig:time&operation} shows the average time cost and operation times (in this case, we recorded the number of mouse clicking during the task). 
It illustrates that \textit{DualMotion} allowed users to design motions more efficiently. Notably, the time cost and the operation times of the free motion design task (Figure~\ref{fig:time&operation} (green)) included the time for envisioning their target motion for each participant. 
Therefore, the participants are able to manage \textit{DualMotion} and create the motion following their intent in a short time. 

\begin{figure}[t]
    \centering
    \includegraphics[width=1\linewidth]{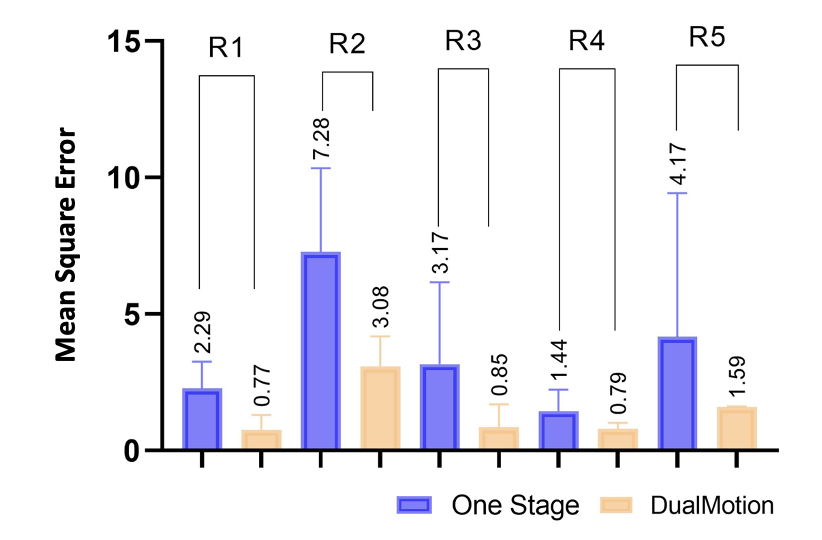}
    \caption{MSE loss results between the user design results and the references. (R1, R2, etc are indexes of the reference motion.)}
    \label{fig:mse}
\end{figure}

We show the post-experiment questionnaire results below. 
The SUS metrics evaluation result is shown in Table~\ref{table:SUS}. 
All of the participants claimed that they felt satisfaction of the overall interface, and they would imagine that most of the people would learn to use this system very quickly. 
Furthermore, 86\% (12/14) of the participants said that they would like to use \textit{DualMotion} frequently to design the prototype of character animations. 
Moreover, 79\% (11/14) and 71\% (10/14) of the participants said that they were confident when using the system and that the functions were well-integrated, respectively. 
\textit{DualMotion} scored 75.6 out of 100 through the SUS metrics, which implies the overall good usability.

\begin{figure}[t]
    \centering
    \includegraphics[width=0.92\linewidth]{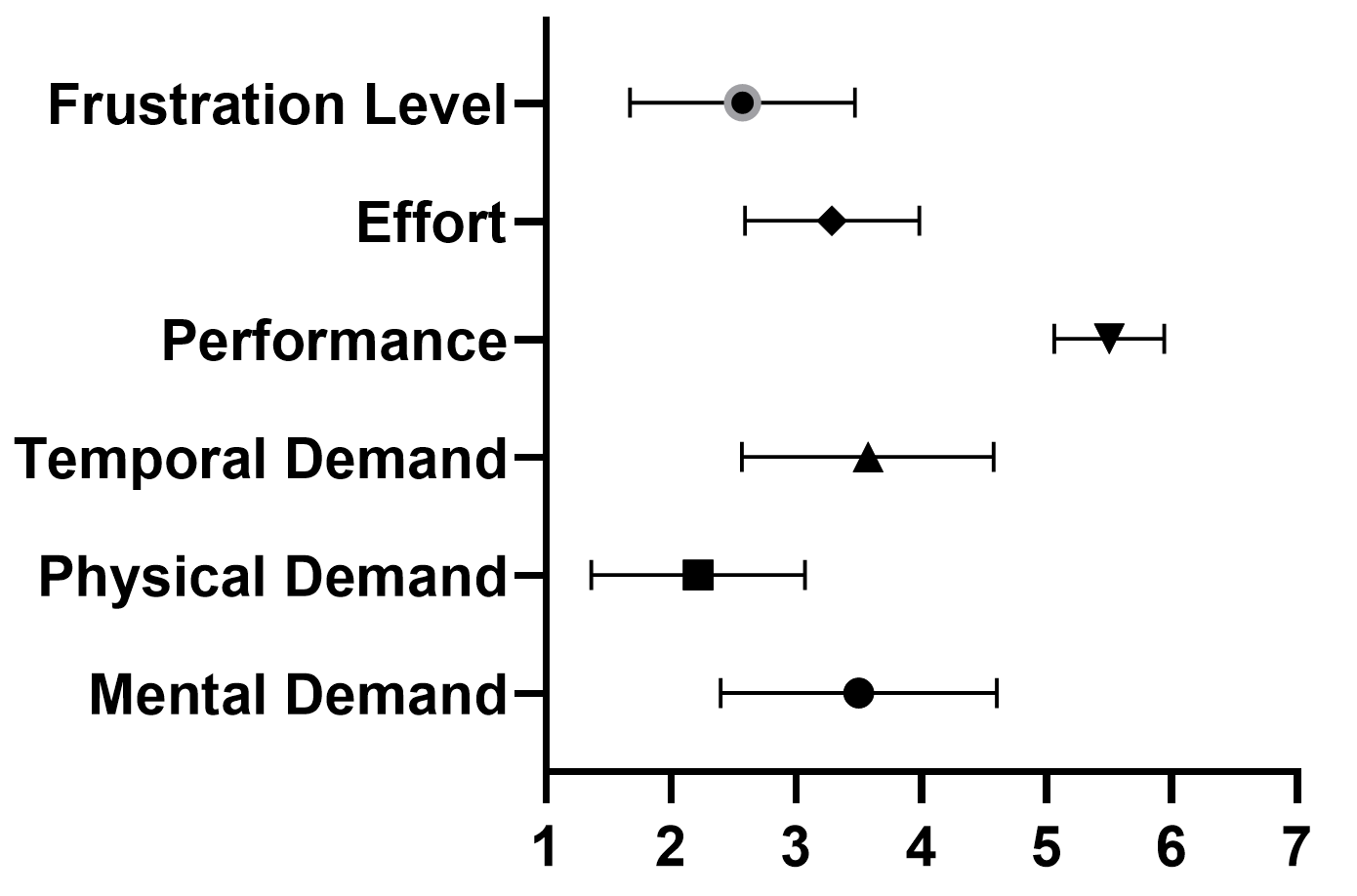}
    \caption{The result of \textit{raw} NASA-TLX evaluation.}
    \label{fig:nasatlx}
\end{figure}

Lastly, the work load evaluation result is shown in Figure~\ref{fig:nasatlx}.
The high score of `\textit{Overall Performance}' illustrates that the participants were satisfied with their own animations. 
By contrast, the low level of '\textit{physical demand}' and '\textit{frustration level}' implies that the large loads of labor unnecessary when utilizing \textit{DualMotion} to design a character animation. 

\begin{table}[t]
  \centering
  \caption{Results of the post-experiment SUS metrics questionnaire. $\big\Uparrow$ indicates higher scores are better. $\big\Downarrow$ for the other case. The total score is 75.6 out of 100.}
  \label{table:SUS}
  \begin{tabular}{l|p{0.62\linewidth}ccc}
    \hline
    \# & Questions  & Mean & SD \\
    \hline \hline
    1 & I think that I would like to use \textit{DualMotion} frequently. $\big\Uparrow$  & 4.00  & 0.48\\
    2 & I found \textit{DualMotion} unnecessarily complex. $\big\Downarrow$  & 2.07 & 0.92\\
    3 & I thought \textit{DualMotion} was easy to use. $\big\Uparrow$  & 3.93  & 0.99\\
    4 & I think that I would need the support of a technical person to be able to use this \textit{DualMotion}. $\big\Downarrow$ &  2.64  & 1.28\\
    5 & I found the various functions in this \textit{DualMotion} were well integrated. $\big\Uparrow$ & 3.86 & 0.66\\
    6 & I thought there was too much inconsistency in \textit{DualMotion}. $\big\Downarrow$ & 1.92 &  0.91\\
    7 & I would imagine that most people would learn to use \textit{DualMotion} very quickly. $\big\Uparrow$ &  4.50  &  0.52\\
    8 & I found \textit{DualMotion} very cumbersome to use. $\big\Downarrow$ &  2.07  &  0.99\\
    9 & I felt very confident using \textit{DualMotion}. $\big\Uparrow$ &  4.07  &  0.73\\
    10 & I needed to learn a lot of things before I could get going with \textit{DualMotion}. $\big\Downarrow$ &  1.42  &  0.51\\
    \hline
  \end{tabular}
\end{table}

\subsection{Animation Prototype}
\begin{figure}[t]
    \centering
    \includegraphics[width=1\linewidth]{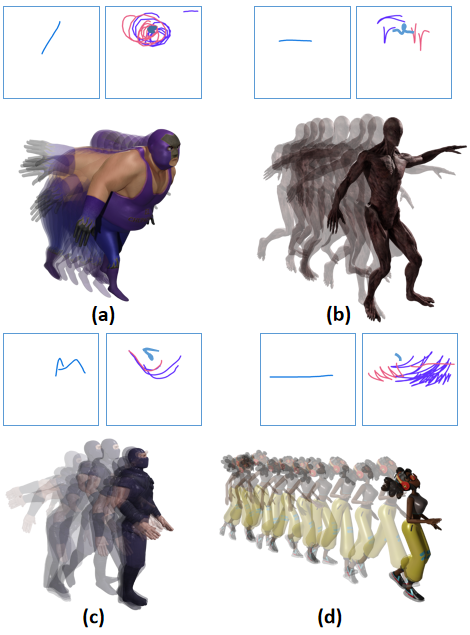}
    \caption{Motion retargeting results on Mixamo's character models. The query sketch is showed above each result which are global editorial strokes (left) and local editorial strokes (right). The different colors in local design stage represent the design history of various joints which are head (cyan), left hand (violet), and right hand (pink) respectively.}
    \label{fig:retarget}
\end{figure}

To evaluate whether the design results are qualified enough to make a animation prototype, we retargetted the users' free designed skeletal motion on to several character models with various body sizes (Michielle, Maynord, Ninja, and Ortiz from \textit{Mixamo}~\cite{mixamo}).
We observed that the animation prototype was consistent with the user's editorial intent.
As shown in Figure~\ref{fig:retarget}, the user editorial intents were \textit{a crash} (a) \textit{a zombie walk} (b), \textit{a standing long jump} (c), and \textit{a run with joy} (d). More results are shown in the supplemental video.
We also conduct oral interviews with users after finishing their free designed animation. Most of the users said that although they only had a rough idea at the very first and not clearly know exactly how to implement movement (e.g., someone wanted to design a zombie walk but did not exactly know how the hands or head move.), they can achieved their goal and were satisfied with the results.
The results verify that \textit{DualMotion} can not only supports high-precision motion design, but can also help users successfully complete their animations with only vague ideas and achieve satisfaction with their own work.

\section{Discussion and Limitation}
\label{sec:7}

\subsection{Applications on Mobile Platforms}
In our user study, we found that \textit{DualMotion} supports novices to create character animations in two usage scenarios:
1) to create a character animation with a clear target (given a reference motion sequence), and
2) to create a character animation without a target.
Particularly, \textit{DualMotion} reduces the operations in casual character animating.
As an extreme case, animating on mobile devices with touch screen strictly limits on the count and accuracy of user operation.
Actually verifying \textit{DualMotion}'s extendibility on tablets and smartphones is one of our future work.

\subsection{Trade-off between Dataset Size and Computation Performance}
The current prototype of \textit{DualMotion} and the user study are based on a light-weight motion database. \textsc{blue}{(55 motion data in total)}
Our system updates the projection matrix every time the camera is relocated by users and then applies on all motion sequences in the database. \textcolor{black}{The average retrieval time is around 0.84 second for each query.}
As the size of the database grows, the computation performance may not be able to meet the demands of real-time interaction sensitively.
\textcolor{black}{To be mentioned that the retrieval task is implemented by a single node loop on CPU, a parallel processing pipeline on GPU is required to increase the efficiency.
Besides,} we will extend the current database and accelerate the algorithm to improve the diversity of design results.

\subsection{User Sketches in Design Process}
The current user-input stroke is made up by uniformly distributed points, which means that the input omits the velocity of user drawing.
We only enable head and hand nodes authoring in the prototype for simplicity of the design process; however it limits the diversity of design results.
Exploration on inputs that contain more information is yet another future work for more professional users.

\subsection{Creativity Support with Two-Stage Scheme}
\textit{DualMotion} adopted a two-stage scheme with both global and local stages \cite{huang2021dualface,sca21}. Along with the perception capabilities of human, designing the target with a bottom-up scheme is difficult. For example, we usually have a gradually clear design intention during the design process, and only an ambiguous target in our mind at the initial step. We thought the two-stage scheme is an approach of human-centered design thinking and could be useful in various creativity support systems, such as illustration and story designs.

\textcolor{black}{As technical limitations, the current prototype of \textit{DualMotion} combined the limbs' motion data to the torso's of the character model. The current implementation may produce the artifacts such as the unnatural body balance. A lightweight algorithm for automatically correcting the full body balance can be considered for further development, such as a double inverted pendulum model~\cite{MIG17}.}

\section{Conclusion}
\label{sec:8}
In this work, we proposed \textit{DualMotion}, a casual motion design UI for character animations.
This helps common users create character motion animation, as they are not always able to conduct accurate and a large amounts of operations.
In this case, editing motion by using professional software, which has many complex functions when the operation count and range are limited, poses further challenges.
The proposed system utilizes a two-stage design scheme with global and local motion retrieval and composition to tackle the challenging task of decomposing full-body motions into lower limb and upper limb movements.
Users are allowed to query the database with simple and rough sketch input and retrieve desirable results by a data-driven manner. Lastly, a user study is conducted, which verified that \textit{DualMotion} is available to supports users to create character animations in low labor demand, even with only vague ideas.

{\small
\bibliographystyle{ieee_fullname}
\bibliography{ref}
}

\end{document}